\documentclass{emulateapj}
\shorttitle{225 GHz Opacity Monitoring at the Greenland Summit}
\shortauthors{Matsushita et al.}

\slugcomment{Accepted on Oct.\ 29, 2016}

\begin{document}

\title{3.5-Year Monitoring of 225 GHz Opacity at the Summit of
	Greenland}

\author{Satoki Matsushita\altaffilmark{1},
%		\email{satoki@asiaa.sinica.edu.tw}
	Keiichi Asada\altaffilmark{1},
    Pierre L. Martin-Cocher\altaffilmark{1},
	Ming-Tang Chen\altaffilmark{1},
	Paul T. P. Ho\altaffilmark{1},
	Makoto Inoue\altaffilmark{1},
	Patrick M. Koch\altaffilmark{1},
	Scott N. Paine\altaffilmark{2},
	David D. Turner\altaffilmark{3}
	}

\altaffiltext{1}{Academia Sinica Institute of Astronomy and
	Astrophysics, P.O.\ Box 23-141, Taipei 10617, Taiwan, R.O.C.}
\altaffiltext{2}{Harvard-Smithsonian Center for Astrophysics,
	60 Garden Street, MS-78, Cambridge, MA 02138, U.S.A.}
\altaffiltext{3}{Global Systems Division, Earth System Research
	Laboratory, National Oceanic and Atmospheric Administration,
	325 Broadway, Boulder, CO, 80305, U.S.A.}
%\altaffiltext{3}{National Severe Storms Laboratory,
%	National Oceanic and Atmospheric Administration,
%	120 David L. Boren Boulevard, Norman, OK, 73072, U.S.A.}

\begin{abstract}
We present the 3.5-yr monitoring results of 225 GHz opacity at the
summit of the Greenland ice sheet (Greenland Summit Camp) at an
altitude of 3200 m using a tipping radiometer.
We chose this site as our submillimeter telescope (Greenland
Telescope; GLT) site, because conditions are expected to have low
submillimeter opacity and because its location offers favorable
baselines to existing submillimeter telescopes for global-scale Very
Long Baseline Interferometry (VLBI).
%There are presently no other millimeter or submillimeter wave
%observatories geographically near this remote and unique site.
%condition and possible to have very long baseline lengths
%with other existing submillimeter telescopes for very long baseline
%interferometry (VLBI) because of no telescope around this remote
%and unique site.
The site shows a clear seasonal variation with the average opacity
lower by a factor of two during winter.
The 25\%, 50\%, and 75\% quartiles of the 225 GHz opacity during the
winter months of November through April are 0.046, 0.060, and 0.080,
respectively.
For the winter quartiles of 25\% and 50\%, the Greenland site is
about $10\%-30\%$ worse than the Atacama Large
Millimeter/submillimeter Array (ALMA) or the South Pole sites.
%, but better in summer than the ALMA site.
Estimated atmospheric transmission spectra in winter season are
similar to the ALMA site at lower frequencies ($<450$ GHz), which are
transparent enough to perform astronomical observations almost all of
the winter time with opacities $<0.5$, but $10\%-25\%$ higher
opacities at higher frequencies ($>450$ GHz) than those at the ALMA
site.
This is due to the lower altitude of the Greenland site and the
resulting higher line wing opacity from pressure-broadened saturated
water lines in addition to higher dry air continuum absorption at
higher frequencies.
Nevertheless, half of the winter time at the Greenland Summit Camp
can be used for astronomical observations at frequencies between
450 GHz and 1000 GHz with opacities $<1.2$, and 10\% of the time
show $>10\%$ transmittance in the THz (1035 GHz, 1350 GHz, and
1500 GHz) windows.
Summer season is good for observations at frequencies lower than
380 GHz.
One major advantage of the Greenland Summit Camp site in winter is
that there is no diurnal variation due to the polar night condition,
and therefore the durations of low-opacity conditions are
significantly longer than at the ALMA site.
Opacities lower than 0.05 or 0.04 can continue for more than 100
hours.
%, and for several tens of hours even for opacities lower than 0.03.
%There are many times when opacities lower than 0.05 exist for periods
%longer than 100 hours, and events where the opacity is lower than
%0.03  for many tens of hours.
Such long stable opacity conditions do not occur as often even at the
South Pole; it happens only for the opacity lower than 0.05.
%, and the opacity lower than 0.03 continues up to only 10 hours.
Since the opacity variation is directly related to the sky
temperature (background) variation, the Greenland Summit Camp is
suitable for astronomical observations that need unusually stable sky
background.
%The Greenland Summit Camp is, therefore, suitable for astronomical
%observations that need unusually stable sky conditions.
\end{abstract}


\section{Introduction}
\label{sect-intro}

As various technologies for submillimeter (submm) wave observations
have advanced, the prospects for very long baseline interferometry
(VLBI) at submm wavelengths have become a reality.
Operations at shorter wavelengths improve proportionately the spatial
resolution as compared with current centimeter and millimeter VLBI
observations; we anticipate that it will be possible to resolve
astronomical sources in greater details, by a factor of 10.

The demand for better angular resolution is quite strong, especially
for the direct imaging of nearby supermassive black holes (SMBHs);
Sagittarius A$^{*}$ (Sgr A$^{*}$), which is located at the center of
our Galaxy and therefore the nearest SMBH, has been imaged at various
wavelengths using the VLBI technique.
The observed size of Sgr A$^{*}$ is obviously affected by the
interstellar scattering at 3 mm or longer wavelengths, following the
$\lambda^{2}$ scattering law \citep[e.g.,][]{she05,bow06}.
This effect, however, lessens at shorter wavelengths, and the size is
observed to deviate from the $\lambda^{2}$ scattering law at 1.3 mm
\citep{doe08}.
This has strongly motivated submm-VLBI observations toward nearby
SMBHs to resolve and image emission from their vicinities.

Upper limits of the intrinsic sizes of SMBHs have been measured so
far for Sgr A$^{*}$ \citep{doe08} and the nucleus of M87
\citep{doe12}; upper limits to the sizes of both sources are about
$40~\mu$arcsec.
These results indicate that much longer baselines and/or higher
frequencies are needed to resolve and image the SMBHs.
We, therefore, started to look for a new site to perform submm-VLBI
observations with substantially longer baselines than before.

\section{Site Selection}
\label{sect-site}

For the site selection, we set criteria as follows:
\begin{enumerate}
\item Annual precipitable water vapor (PWV) of less than 3 mm for low
	submm opacity.
\item Longest possible baselines with existing submm telescopes, for
	obtaining the highest angular resolution (this also means that we
	do not consider sites that already have submm telescopes).
\item Overlapping sky coverage with the Atacama Large
	Millimeter/submillimeter Array (ALMA) to achieve the highest
	possible sensitivity.
\item Accessibility to the site for maintenance and operations.
\end{enumerate}

%%% Write more details about the site selection criteria?

We checked the satellite-based PWV data measured by the Moderate
Resolution Imaging Spectroradiometer (MODIS) on the NASA Aqua and
Terra satellites for potential sites with respect to the locations of
the available submm telescopes.
We found three potential broad regions of interest; western China and
Tibet, the highest mountains of southern Alaska, and the high Arctic
polar desert, including northern Canada and Greenland.
The Western China and Tibet regions do not have common sky with ALMA,
so they do not meet the criterion (3).
The tallest peaks in Alaska (e.g., Denali, or former official name
Mount McKinley) are protected or otherwise inaccessible, so they do
not meet the criterion (4).
The summit of the Greenland ice sheet, on the other hand, has low
PWV conditions throughout the year, has a common sky coverage with
ALMA, will create the longest baseline length of about 9000 km for
the submm-VLBI, and already has a research facility, Summit Camp,
which is operated by CH2M Hill\footnotemark[1].
\footnotetext[1]{\url{https://www.ch2m.com/}} Polar Services (CPS)
for the U.S.\ National Science Foundation (NSF).
The Summit Camp, therefore, meets all four criteria.

The Greenland Summit Camp is located at $72\fdg57$ N latitude and
$38\fdg46$ W longitude, at an altitude of 3200 m.
The temperature is very low, with winter temperatures between
$-30\arcdeg$C and $-60\arcdeg$C, and summer temperatures between
$0\arcdeg$C and $-30\arcdeg$C \citep{lau10,mar14}.
Due to the combination of the high altitude and the low temperature,
very low opacity is expected.
Furthermore, the NSF is currently funding the Integrated
Characterization of Energy, Clouds, Atmospheric state, and
Precipitation (ICECAPS) project at the Summit Camp, which is using
active and passive ground-based remote sensors, including two
radiometers that observe 16 frequencies from 22.2 GHz to 150.0 GHz,
to provide the first complete description of cloud properties above
this site \citep{shu13}.
They determine the annual cycle of PWV using the radiometer
observations at the central frequencies of 23.8, 31.4, 90, and
150 GHz as well as from radiosondes that are launched twice daily by
ICECAPS technicians.
We, therefore, decided to put a 225 GHz tipping radiometer at this
site to measure the atmospheric opacity conditions for possible submm
VLBI operations (see the next section for the reason to choose the
measurement frequency).

Our group is in the process of deploying a 12 m diameter submm
telescope to Greenland.
% the future submm astronomy and astrophysics observations.
Overall explanations of this project, the Greenland Telescope (GLT)
project, are described in \citet{ino14} for submm-VLBI science and
technical details, and in \citet{hir16} for single-dish science
cases.
In addition, more detailed information about antennas, receivers, and
software for the GLT project are in \citet{raf16}, \citet{gri14}, and
\citet{pat16}, respectively.

\section{Measurement and Data Reduction}
\label{sect-meas}

For the opacity measurement at the summit of the Greenland ice sheet,
we procured a 225 GHz tipping radiometer, RPG-225 Radiometer, from
Radiometer Physics GmbH (RPG).
This instrument has the ability to operate in very cold environments,
and indeed some units are operating at Arctic and Antarctic areas
(two radiometers operating as part of ICECAPS mentioned above are
also from RPG).
The reason for the choice of this operating frequency is that there
are many site survey results from all over the world, including the
current submm telescope sites, such as the summit of Mauna Kea, the
ALMA (Chajnantor and Pampa la Bola) site, and South Pole.
The radiometer has an uncooled double side band heterodyne receiver
with a bandwidth of 1 GHz.
A tipping paraboloid mirror, which can rotate $360\arcdeg$ with its
half power beam width of $0\fdg5$, is installed in front of the feed
horn.
A Gortex window covers $\pm90\arcdeg$ from zenith to allow sampling
of the sky signal, and a black body target, whose temperature is
monitored by a thermometer, is located at the bottom (i.e.,
$180\arcdeg$ from zenith).
A 140 W blower is located below the window and provides heating to
prevent ice formation and accumulation on the window.

%%% Put the radiometer picture on MSF.
\begin{figure}[t]
%\epsscale{0.85}
\plotone{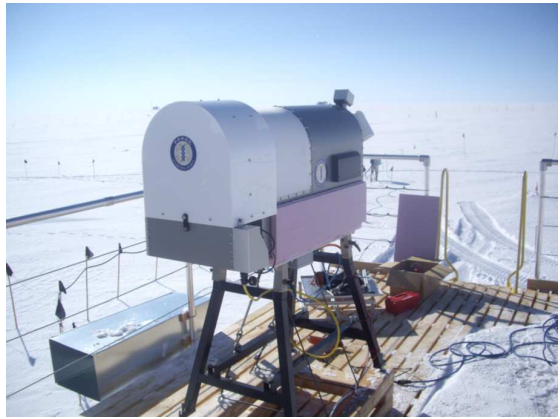}
%\begin{center}
%\includegraphics%[angle=90,width=0.9\linewidth]{IMGP2572.jpg}
%\end{center}
\caption{The 225 GHz tipping radiometer located on the roof of the
	Mobile Science Facitily (MSF) at the Greenland Summit Camp.
\label{fig-rpg}}
\end{figure}

We obtained the radiometer in the autumn of 2010, deployed it on the
roof of our institute in Taipei, and conducted functional and gain
stability tests.
We then moved the radiometer to the summit of Mauna Kea, Hawaii, in
the end of 2010, to check the consistency with the 225 GHz tipping
radiometer at the Caltech Submillimeter Observatory (CSO).
We put our radiometer near the CSO with the same tipping direction,
measured the opacity for about 2 weeks, and confirmed that the
results were consistent with each other (linear regression
coefficient = 1.04).
After this, we moved the radiometer to the Polar Environment
Atmospheric Research Laboratory (PEARL), located on a ridge at an
altitude of 610 m at $80\fdg05$ N and $86\fdg42$ W, 15 km away from
the Eureka weather station on Ellesmere Island, Canada.
We measured the atmospheric opacity for 3 months between late winter
to early spring, and the results are reported in \citet{asa12} and
\citet{mat13}.
After this measurement, we moved the radiometer to the Greenland
Summit Camp.
The radiometer was installed on the roof of the Mobile Science
Facility \citep[MSF;][]{shu13}, which was built by CPS in support of
the ICECAPS project.
%is operated under the auspices
%of the US Department of Energy Atmospheric Radiation Measurement
%(ARM) program \citep{ack03}.
The roof deck is at about 3 m above the snow surface
(Fig.~\ref{fig-rpg}).
The measurement was started from August 17th, 2011, and data were
collected until February 12th, 2015.
%temporally finished on February 12th, 2015, due to an
%instrumentation problem.
We present here the data for these $\sim3.5$ years.

To measure the atmospheric opacity, we adopt the tipping method;
we observe five angles from horizon ($90\arcdeg$ = zenith,
$42\arcdeg$, $30\arcdeg$, $24\arcdeg$, and $19\fdg2$, which
corresponds to $\sec(z)$ of 1.0, 1.5, 2.0, 2.5, and 3.0) with
4 second integration at each angle.
We scan the mirror from south to north, namely observe the five
angles in both the southern and the northern sky (for the
measurements at Greenland).
The black body target in the bottom of the radiometer is observed
before and after the scan for the gain calibration.
The total duration of a tipping measurement is 75 seconds, and each
tipping is performed every 10 minutes.
Between the tipping measurements, the mirror is pointed toward the
zenith and the sky data are recorded every 1 second.
The output voltages, together with the mirror position, the black
body target temperature, and other monitoring data, are recorded
every 1 second into the hard disk of the host computer in binary
format.
The raw binary data are downloaded from Greenland to Taiwan regularly
via internet.

\begin{figure}[t]
%\epsscale{0.85}
\begin{center}
\includegraphics[angle=90,width=0.9\linewidth]
	{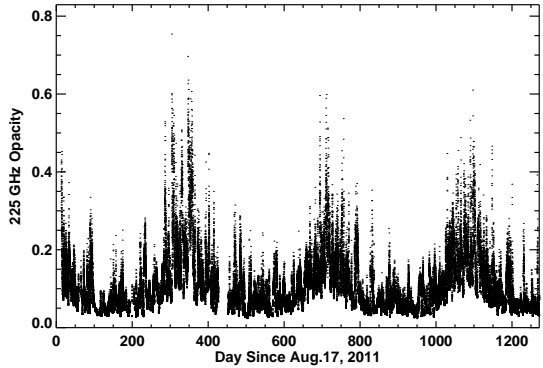}
\end{center}
\caption{Time variation plot of 225 GHz opacity at the summit of
	Greenland ice sheet.
	The measurement has been started from Aug.~17, 2011, which is
	defined as the day 1 in this diagram.
\label{fig-time}}
\end{figure}

\begin{figure}[t]
%\epsscale{0.85}
\plotone{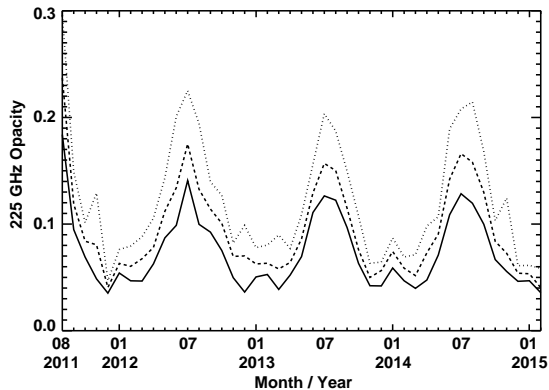}
\caption{Monthly quartile variation of 225 GHz opacity.
	Solid, dashed, and dotted lines are 25\%, 50\%, and 75\%
	quartiles, respectively.
	The first and the last months of this measurement, namely
	Aug.\ 2011 and Feb.\ 2015, have fewer data than the other months,
	so that the statistical significance is low.
\label{fig-month}}
\end{figure}

The tipping data have been reduced with the typical data reduction
method for tipping measurements \citep[e.g.,][]{mat98}.
Since the MSF is a mobile facility on the snow, it can shake due to
wind or human activities inside the facility, so that the radiometer
tipping angle may be affected.
% by this shaking.
In addition, the leveling accuracy for the radiometer is limited, so
that the radiometer may also have a small constant tilt.
We assume that the opacity is the same between the southern and
northern skies, and the small difference in opacities derived from
the southern and northern sky tippings is considered as the result of
the tilt of MSF and/or the radiometer.
Based on this assumption, we calculate the tilt angle and correct for
it when deriving the opacity.
%In Fig.~\ref{fig-tilt}, the derived MSF tilt as a function of time is
%plotted.
%As can be seen, the tilt angle is well within $2\arcdeg$.
In case the difference of the opacities between the southern and
northern skies is large (i.e., when the tilt angle is calculated to
be larger than two degrees; this value is also to allow some
tolerance for the different opacities between the northern and
southern sky),
%, since it is
%difficult to have exactly the same opacities between opposite
%directions of the sky), 
we judge this is due to real opacity differences, and we flag the
data.

\section{Results and Discussions}
\label{sect-res}

\subsection{Time and Seasonal Variations}
\label{sect-res-time}

Fig.~\ref{fig-time} displays the measured time variation of the 225
GHz opacity for the 3.5-year period at the Greenland Summit Camp.
It is clear that there is a seasonal variation; opacity in winter is
low, but high in summer.
Fig.~\ref{fig-month} shows the monthly quartile variations (solid,
dashed, and dotted lines are 25\%, 50\%, and 75\% quartiles,
respectively).
December and March tend to have the best opacity conditions, and July
tends to have the worst.
Note that the first and the last months of this measurement (i.e.,
Aug.\ 2011 and Feb.\ 2015) have fewer data than the other months, so
the statistical significance is lower.
In both diagrams, there is no significant annual difference.
%, but seems better in recent years.

\subsection{Cumulative Distribution and Histogram}
\label{sect-res-cdh}

\begin{figure}[t]
%\epsscale{0.85}
\plotone{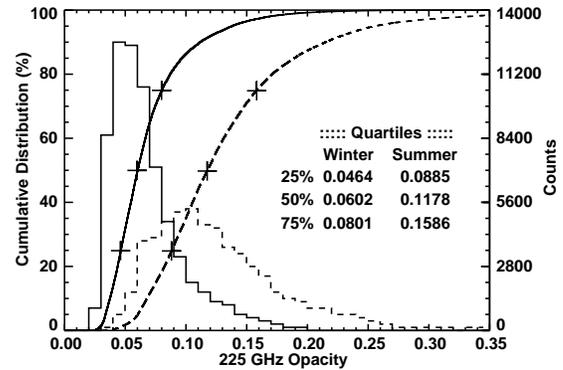}
\caption{Cumulative distribution plots and histograms of 225 GHz
	opacity in winter (solid lines) and summer (dashed lines).
	The vertical axis on the left-hand side is for the cumulative
	distribution plots, and that on the right-hand side is for the
	histograms.
	Crosses on the cumulative distribution plots are the opacity
	quartiles of each season.
	The quartile for winter and summer are also listed in the figure.
\label{fig-cdp}}
\end{figure}
%%% <== Put points at quartiles, rather than long ticks at the bottom?

We also made cumulative distribution plots and histograms of the
measured 225 GHz opacity while separating the seasons into winter and
summer (Fig.~\ref{fig-cdp}).
Here we define winter as between the beginning of November and the
end of April (solid line plots), and summer as May through October
(dashed line plots).
The quartiles for each season are 0.046, 0.060, and 0.080 for 25\%,
50\%, and 75\%, respectively, in winter, and 0.089, 0.118, and 0.159
for summer (see also the crosses and the values in the figure).
It is obvious that the opacity in winter is about half of that in
summer at all the quartiles.
The histograms show that the opacities of 0.04 and 0.10 are the
opacities that occur most often in winter and summer, respectively.

%\section{Comparison with Other Submillimeter Sites}
%\label{sect-comp}

%Using our 3.5 years of 225 GHz opacity statistics at the Greenland
%Summit Camp,
We then compared the opacity quartiles using our 3.5-year period
statistics with those at the ALMA and the South Pole sites, which are
well-established sites for submm observations.
The 225 GHz opacity data for the ALMA site have been obtained from
\citet{rad00} and \citet{rad11}, whose measurements have been made
between April 1995 and April 2006 ($\sim11$ years), and that for the
South Pole site from \citet{cha94,cha95} measured between January and
December 1992 (1 year).
%the NRAO ALMA Site Characterization Homepage\footnote{
%\url{http://science.nrao.edu/alma/site-characterization.shtml}} and
%Simon Radford's Chajnantor Site Evaluation Homepage\footnote{
%\url{http://www.submm.caltech.edu/~sradford/site-eval/}}.
%There are data between April 1995 and April 2006, and we used all the
%data available in these homepages.
%The data for the South Pole site have also been taken from the
%Simon Radford's Chajnantor Site Evaluation Homepage.
%For this site, it only exist 1992 data, and we used these data for
%the analysis.
%All these ALMA and South Pole opacity data have been published in
%the past \citep{rad00,rad11,cha94,cha95}.
Since both the ALMA and the South Pole sites are located in the
southern hemisphere, we define winter as between the beginning of May
through the end of October, and summer as November through April.
The calculated quartiles for these three sites are listed in
Table~\ref{tab-quartile}.

%%% 			25%			50%			75%
%%% GLT:
%%%   Winter:  0.0464451	0.0602593	0.0801158
%%%   am PWV:  0.506161		0.790332	1.2054
%%%   Summer:  0.0884677	0.117822	0.158621
%%% ALMA:
%%%   Winter:  0.0345574	0.0504359	0.0795721
%%%   Summer:  0.0708959	0.131414	0.260993
%%% South Pole:
%%%   Winter:  0.0410000	0.0480000	0.0570000
%%%   Summer:  0.0500000	0.0620000	0.0760000

%%% 			1%			2%			3%			5%			10%
%%% GLT
%%%   Winter:  0.0291581	0.0311075	0.0322820	0.0338087	0.0374949
%%%   am PWV:  0.146877		0.185715	0.211074	0.243827	0.321416
%%%   Summer:  										0.0605747	0.0679466

\begin{deluxetable}{ccccc}
%\begin{deluxetable*}{cccc}
%\tabletypesize{\scriptsize}
\tablecaption{Comparison of 225 GHz opacity quartiles between three
	sites.
	\label{tab-quartile}}
\tablehead{
	\colhead{Site}
		& \colhead{Season}
		& \multicolumn{3}{c}{Quartiles} \\
	& & \colhead{25\%}
		& \colhead{50\%}
		& \colhead{75\%}
	}
\startdata
GL Summit Camp & Winter & 0.046 & 0.060 & 0.080 \\
               & Summer & 0.089 & 0.118 & 0.159 \\
ALMA           & Winter & 0.035 & 0.050 & 0.080 \\
               & Summer & 0.071 & 0.131 & 0.261 \\
South Pole     & Winter & 0.041 & 0.048 & 0.057 \\
               & Summer & 0.050 & 0.062 & 0.076
\enddata
\end{deluxetable}
%\end{deluxetable*}

For winter, the ALMA site is the best at the 25\% quartile, South
Pole is next, and Greenland Summit Camp is the worst, but only by
$\sim0.005$ ($\sim12\%$) difference in opacity between each site.
At the 50\% quartile, the ALMA and South Pole sites are almost the
same, and the opacity is about 0.01 (about 25\%) worse at Greenland
Summit Camp.
At the 75\% quartile, South Pole is the best (40\% better) and the
opacity is the same between the ALMA and the Greenland sites.

\begin{figure*}[t]
\plotone{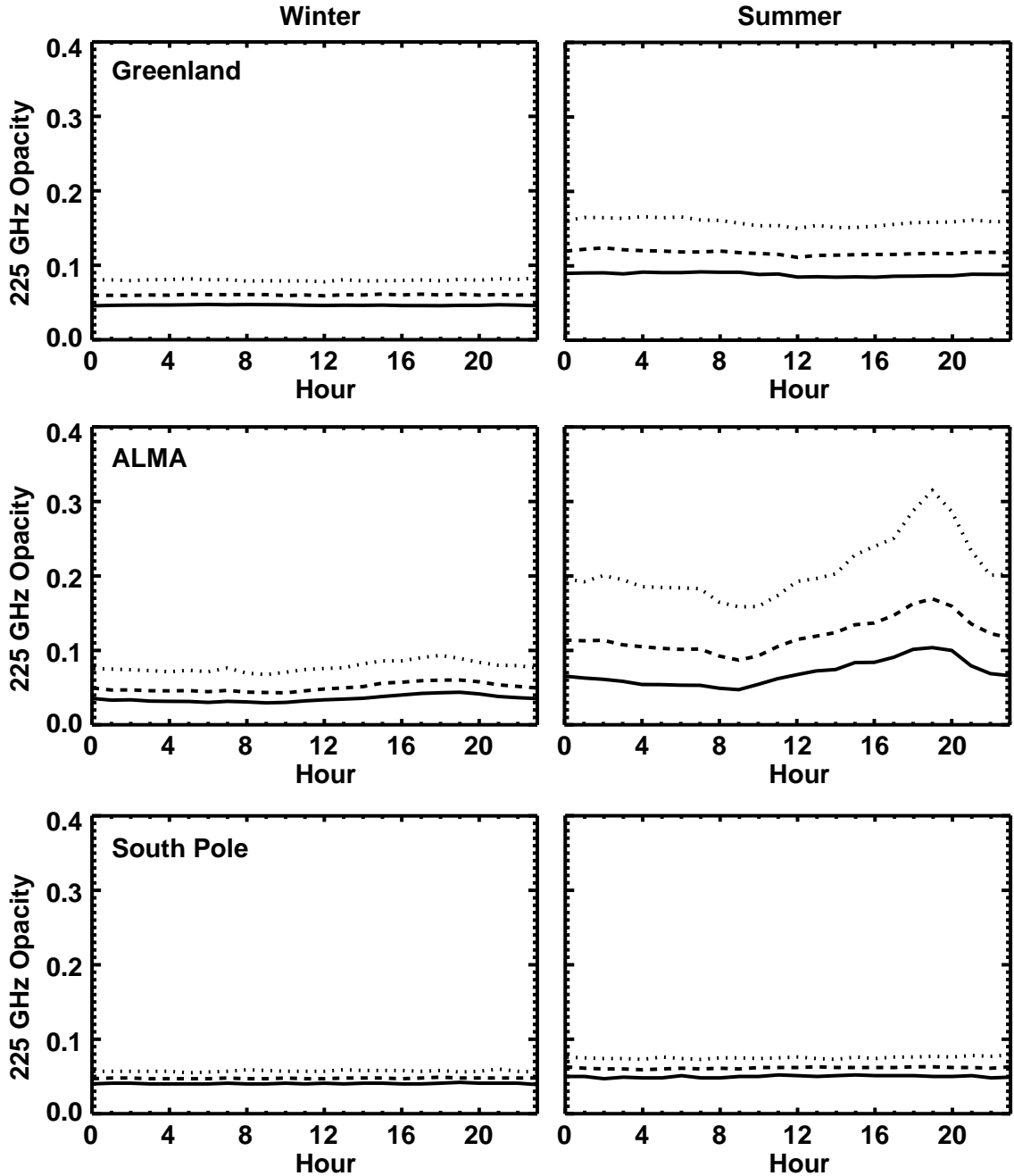}
%\includegraphics[angle=90,width=0.45\linewidth]
%	{diurnal-tip_winter1111-1502.eps}
%\includegraphics[angle=90,width=0.45\linewidth]
%	{diurnal-tip_summer1108-1410.eps}
%%% => Need to match the y-axis range with ALMA/SP.
%\includegraphics[angle=90,width=0.45\linewidth]
%	{diurnal_winter9505-0510_ALMA.eps}
%\includegraphics[angle=90,width=0.45\linewidth]
%	{diurnal_summer9511-0604_ALMA.eps}
%\includegraphics[angle=90,width=0.45\linewidth]{diurnal_winter1992_SP.eps}
%\hspace{13ex}
%\includegraphics[angle=90,width=0.45\linewidth]{diurnal_summer1992_SP.eps}
\caption{Diurnal 225 GHz opacity in winter (left column) and in
	summer (right column) for the Greenland Summit Camp (top row),
	the ALMA site (middle row), and the South Pole site (bottom row).
	Solid, dashed, and dotted lines are 25\%, 50\%, and 75\% of the
	hourly quartiles.
\label{fig-diurnal}}
\end{figure*}

For summer, South Pole is the best for the 25\% quartile, ALMA is the
next, and Greenland is the worst with $\sim0.02$ difference in
opacity between each site.
At the 50\% and 75\% quartiles, South Pole is again the best, but
Greenland is next and the ALMA site is the worst.
The South Pole site is about twice better in opacity as compared with
the Greenland site, and the ALMA site is significantly worse than
these two sites.

%%% Mention about Mauna Kea
We note here that there are many studies that compare the opacity
conditions at the summit of Mauna Kea, which is also a
well-established submm site, with those at the ALMA and South Pole
sites; opacity quartiles of Mauna Kea are $\sim50\%$ higher than
those of the ALMA site \citep{mat99,rad00,rad11,rad16}.

In summary, the South Pole site has little seasonal differences in
opacity over the annual cycle, 
% is very stable over the year
with a factor of two difference between seasons.
On the other hand, the ALMA site has large variations between winter
and summer, with significantly worse conditions in summer, known as
the Bolivian Winter around February and March.
The opacity conditions at the Greenland Summit Camp are roughly
intermediate between the South Pole and the ALMA sites.
%located around in the middle between these two sites.

\subsection{Diurnal Variation}
\label{sect-res-diurnal}

The top row of Fig.~\ref{fig-diurnal} shows the diurnal opacity
variations in winter (left column) and summer (right column) at the
Greenland Summit Camp.
Solid, dashed, and dotted lines are 25\%, 50\%, and 75\% of the
hourly quartiles.
It is obvious that there is no diurnal variation in both seasons.
This can be easily explained by the polar conditions; only nighttime
in winter and daytime in summer.
It is also clear that the opacity in winter is half of that in
summer, as mentioned in the cumulative distribution plot
(Fig.~\ref{fig-cdp}) above.
%Variation is small in winter.

The middle and bottom rows of Fig.~\ref{fig-diurnal} show the diurnal
variation at the ALMA and the South Pole sites, respectively.
It is obvious that there is a clear diurnal variation in the ALMA
data, which is naturally explained by the mid-latitude conditions,
leading to a strong diurnal cycle compared to polar regions, while
the South Pole data are very similar to the Greenland Summit Camp
(i.e., no diurnal variation).

For winter, the South Pole site is always the best at all quartiles.
Opacities at the 25\% and 50\% quartiles at the ALMA site are almost
always statistically better than those at the Greenland Summit Camp,
but for the 75\% quartile, the daytime opacity is statistically
better at the Greenland Summit Camp than at the ALMA site due to the
diurnal variation.

\begin{deluxetable*}{cccccccccccccc}
\tabletypesize{\scriptsize}
\tablecaption{Atmospheric opacity counts per year over specific time
	intervals.
	%Statistics (counts per year) of opacity conditions
	%continuously lower than 0.05, 0.04, and 0.03 for durations longer
	%than 24, 50, 100, 150, and 200 hours.
	\label{tab-duration}}
\tablehead{
	\colhead{Site}
		& \multicolumn{5}{c}{$\tau_{\rm 225 GHz} < 0.05$}
		& & \multicolumn{4}{c}{$\tau_{\rm 225 GHz} < 0.04$}
		& & \multicolumn{2}{c}{$\tau_{\rm 225 GHz} < 0.03$} \\
	& \colhead{$>24$ h}
		& \colhead{$>50$ h}
		& \colhead{$>100$ h}
		& \colhead{$>150$ h}
		& \colhead{$>200$ h}
	& & \colhead{$>24$ h}
		& \colhead{$>50$ h}
		& \colhead{$>100$ h}
		& \colhead{$>150$ h}
		%& \colhead{$>200$ h}
	& & \colhead{$>24$ h}
		& \colhead{$>50$ h} \\
		%& \colhead{$>100$ h}
		%& \colhead{$>150$ h}
		%& \colhead{$>200$ h} \\
	& \multicolumn{5}{c}{Counts / yr}
		& & \multicolumn{4}{c}{Counts / yr}
		& & \multicolumn{2}{c}{Counts / yr}
	}
\startdata
GL Summit Camp & 17.7 &  6.4 & 2.0 & 1.2 & 0.8
               & &  6.0 &  2.4 & 1.2 & 0.4 %& 0.0
               & &  1.2 &  0.4 \\ %& 0.0 & 0.0 & 0.0 \\
ALMA           & 30.6 & 10.7 & 1.5 & 0.2 & 0.0
               & & 17.6 &  4.7 & 0.0 & 0.0 %& 0.0
               & &  4.1 &  0.3 \\ %& 0.0 & 0.0 & 0.0 \\
South Pole     & 32.6 & 13.6 & 5.4 & 2.7 & 2.7
               & &  8.1 &  2.7 & 0.0 & 0.0 %& 0.0
               & &  0.0 &  0.0 %& 0.0 & 0.0 & 0.0
\enddata
\end{deluxetable*}

\begin{figure}[t]
%\epsscale{0.85}
\plotone{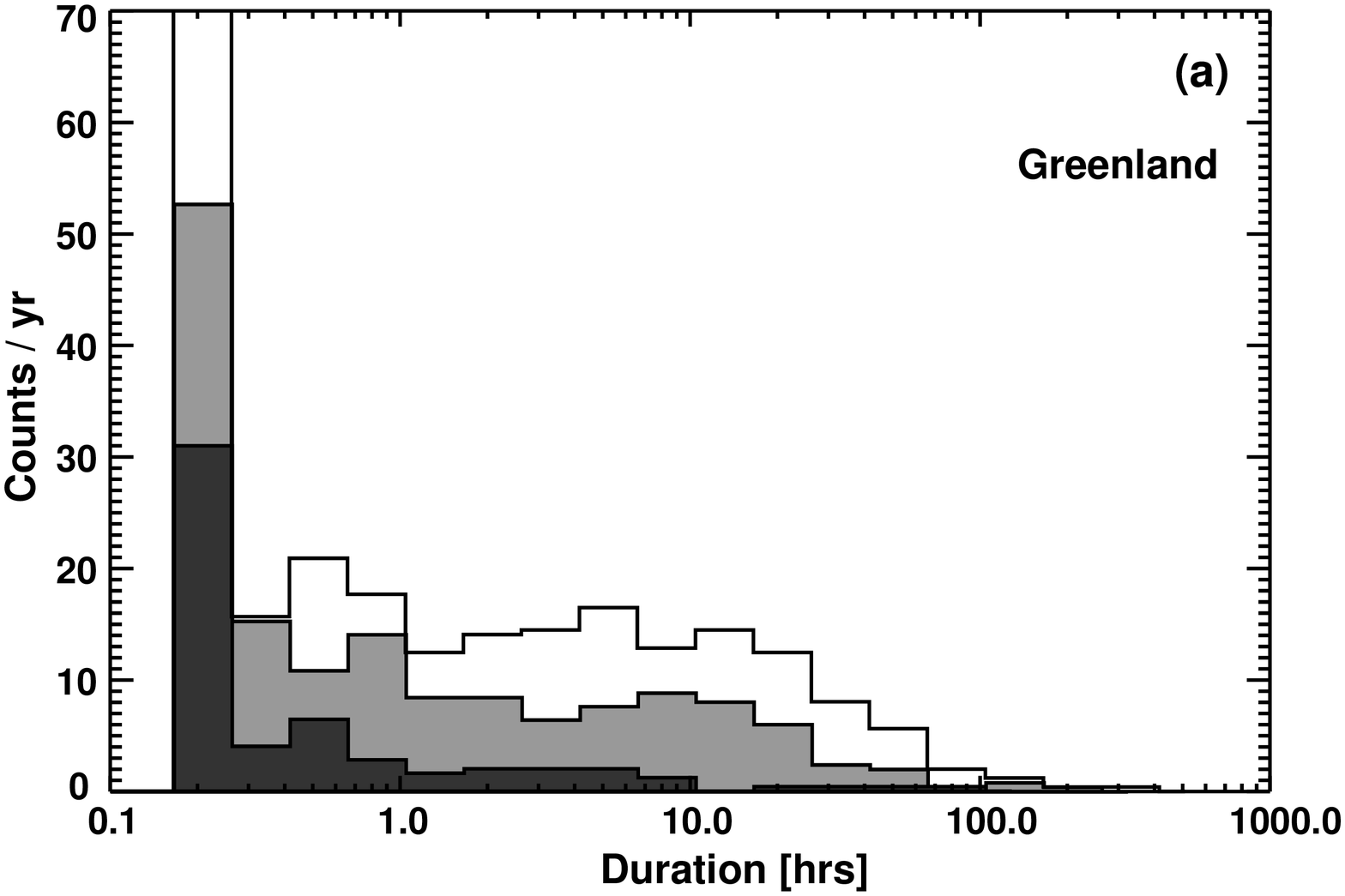}
\plotone{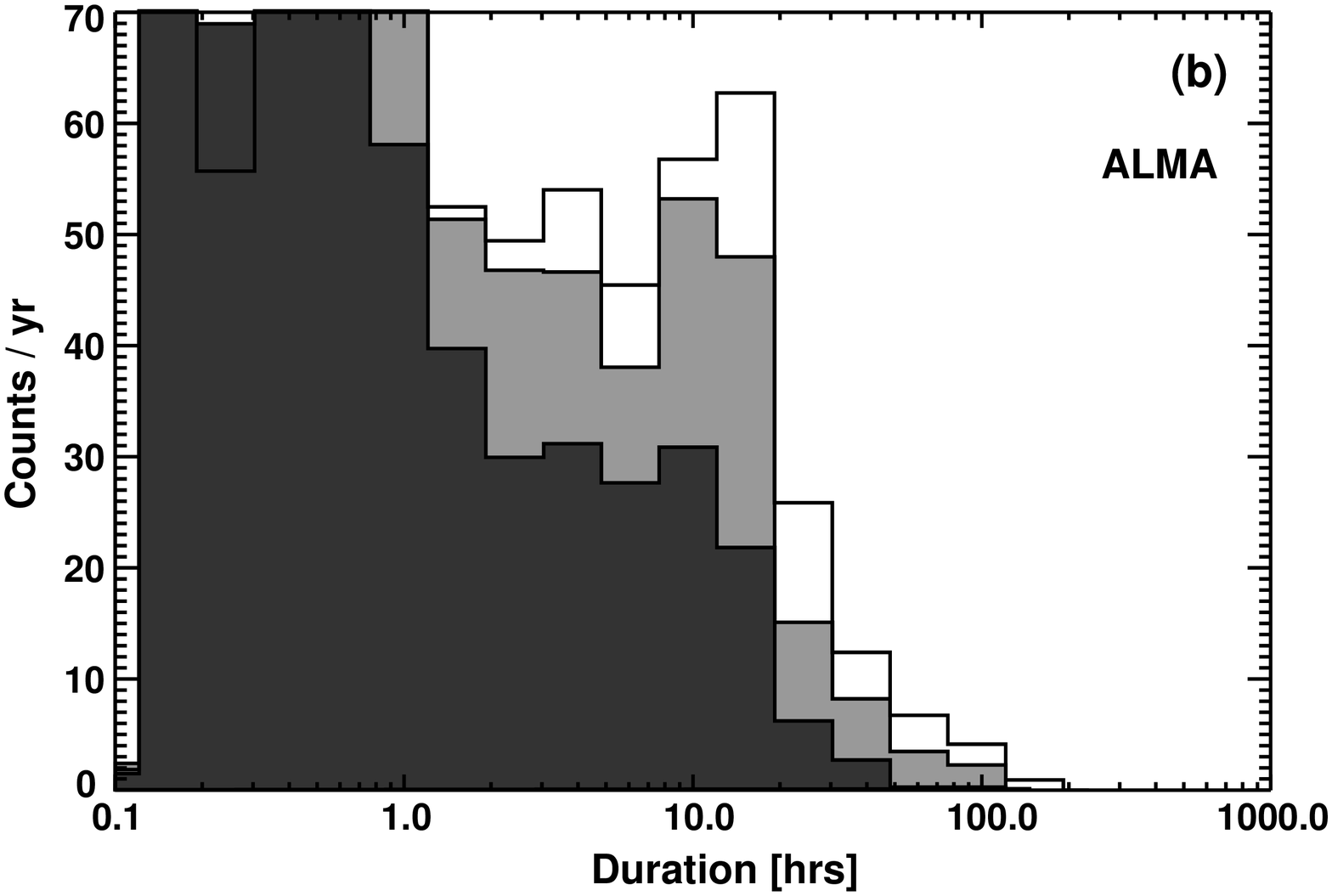}
\plotone{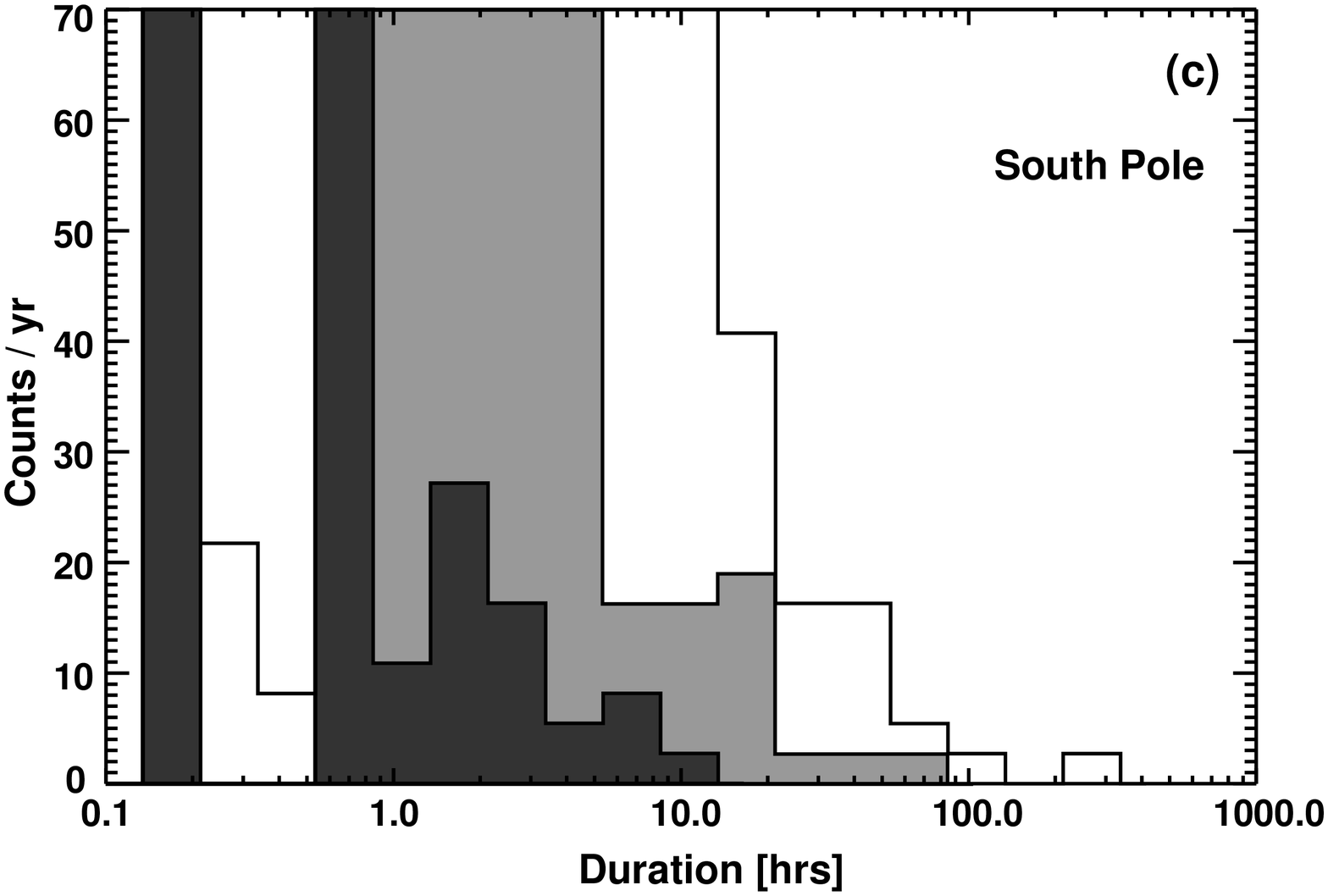}
\caption{Histogram of time duration for opacity continuously lower
	than 0.05 (white), 0.04 (light grey), and 0.03 (dark grey) for
	(a) the Greenland Summit Camp,
	(b) the ALMA site, and
	(c) the South Pole site.
	Vertical axis is counts per year.
\label{fig-duration-hist}}
\end{figure}

For summer, the South Pole site is again always the best at all
quartiles.
For opacities at the 25\% and 50\% quartiles, the Greenland Summit
Camp is always better between $\sim12$ and $\sim20$ hour than the
ALMA site (i.e., daytime at the ALMA site) due to no diurnal
variation.
For opacities at the 75\% quartile, the Greenland Summit Camp is
always better than the ALMA site.
%at whatever time in a day statistically.

%For 25\% and 50\% quartiles in winter, diurnal opacities are almost
%always statistically worse at the Greenland Summit Camp than the ALMA
%site, but for 75\% quartile in winter, the daytime opacity is
%statistically better at the Greenland Summit Camp than the ALMA site
%due to the diurnal variation.
%For 25\% and 50\% quartiles in summer, daytime opacity is always
%better in the Greenland Summit Camp than that in the ALMA site due to
%no diurnal variation.
%For 75\% quartile in summer, on the other hand, the opacity at the
%Geenland Summit Camp is always statistically better than that at the
%ALMA site.
%For the South Pole site, opacity condition at the Greenland Summit
%Camp is always statistically worse than that at the South Pole site,
%but the opacity condition in winter is closer to that at the South
%Pole site than the opacity difference in summer.

\subsection{Duration of Opacity Lower than Certain Values}
\label{sect-res-duration}

%%% Need to update based on the new figures.
We then calculated time durations of opacity conditions continuously
lower than certain values.
We focused on opacities lower than 0.05, 0.04, and 0.03, which are
excellent opacity conditions that only occur in $\lesssim30\%$ of the
winter season at the Greenland Summit Camp.
The resultant histograms for the Greenland Summit Camp are presented
in Fig.~\ref{fig-duration-hist}(a).
%The vertical axis is counts per year to compare with other sites.
Since the tipping measurements are done every 10 minutes as mentioned
in Sect.~\ref{sect-meas}, the lower limit of the measurements are
located at 0.17 hour.
The counts (vertical axis) are normalized with the number of annual
data points, assuming one data point takes 10 minutes
(Sect.~\ref{sect-meas}).
We intentionally cut the count limit to 70 in the plot for
presentation purpose.
%, so that there are more data points above 70
%counts in short durations.
%for short time durations (shorter than 1 hour).

For opacity less than 0.05 or 0.04 at the Greenland Summit Camp,
there were several occasions for which more than 100 hours were
continuously showing opacities lower than those values, and there are
many occasions that continued for more than 10 hours or several
hours.
The counts for durations between 1 hour and 20 hours are almost the
same, and that for durations between 20 hours to several tens of
hours are roughly half of the shorter duration.
For the opacity less than 0.03, there were several occasions when the
opacity was continuously low for more than 10 hours of time, and more
occasions exist for durations of more than an hour.
This is obviously due to the polar conditions (no diurnal variation)
as mentioned above, and it is very difficult to achieve similar
values at other submillimeter sites that are not at polar regions,
such as in the Hawaii (Mauna Kea) or Northern Chile (ALMA) site.

Indeed, we also calculated the time duration of opacity conditions
for the ALMA site under the same opacity conditions as the Greenland
Summit Camp (Fig.~\ref{fig-duration-hist}b), and the difference is
obvious.
At the ALMA site, there are many occasions for the opacity less than
0.05, 0.04, and even 0.03 continues for up to 20 hours, but the
occasions for the duration longer than 20 hours are significantly
smaller, about one-fifth, than the shorter durations.
The duration longer than 100 hours has only been recorded for opacity
lower than 0.05, and never happened for opacities lower than 0.04 or
0.03 in the 11 yr long data.

For the South Pole site (Fig.~\ref{fig-duration-hist}c), the duration
is obviously shorter than that of Greenland Summit Camp.
There are many occasions for the opacity less than 0.05, which
continues up to 10 hours, but there are significantly fewer occasions
for the duration longer than 10 hours.
Similar to the ALMA site, the duration longer than 100 hours has only
been recorded in opacity lower than 0.05, and never happened for
opacity lower than 0.04 or 0.03.
For the opacity lower than 0.03, the duration is only up to about
10 hours, and has never been longer.
Although the South Pole data are taken only for a year and the
opacity statistics are better than for the Greenland Summit Camp
(Sect.~\ref{sect-res-cdh}, \ref{sect-res-diurnal}), good opacity
duration time is shorter.

Table~\ref{tab-duration} shows that low opacity conditions continue
for more than 24, 50, 100, 150, and 200 hours at the Greenland Summit
Camp, ALMA, and the South Pole.
The ALMA and South Pole sites have better atmospheric opacities over
time durations longer than 24 hours and 50 hours at opacity
conditions lower than 0.05 and 0.04, respectively.
But for time durations longer than 100 hours at opacity conditions
lower than 0.04, the Greenland Summit Camp clearly has lower
atmospheric opacities.

\begin{figure}[t]
%\epsscale{0.85}
\plotone{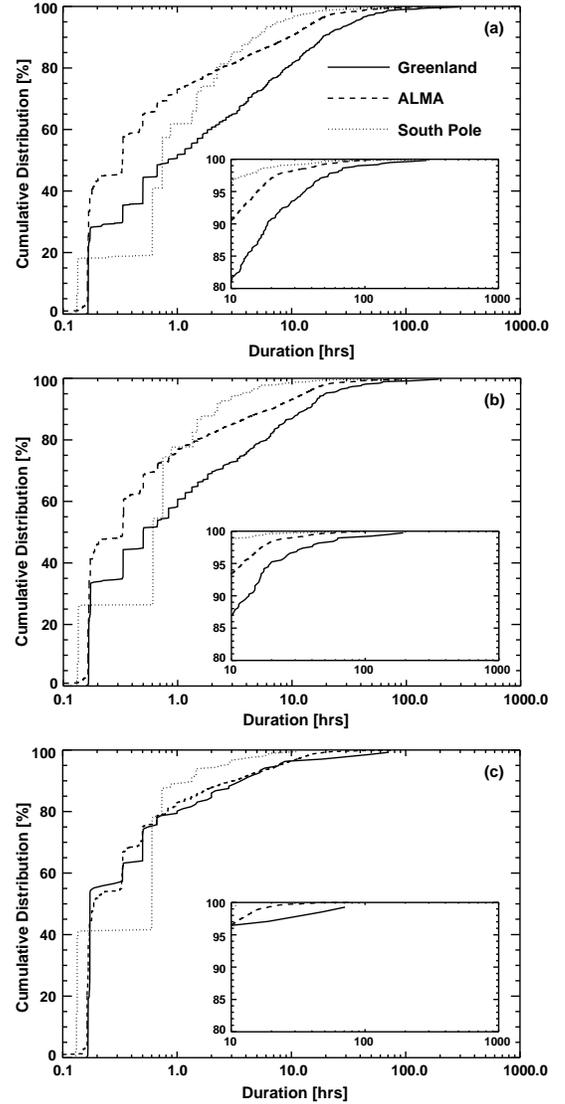}
%\plotone{duration_cumul_tau005.eps}
%\plotone{duration_cumul_tau004.eps}
%\plotone{duration_cumul_tau003.eps}
\caption{Cumulative distribution plots of time duration for opacity
	continuously lower than
	(a) 0.05,
	(b) 0.04, and
	(c) 0.03
	for the Greenland Summit Camp (solid lines), the ALMA site
	(dashed lines), and the South Pole site (dotted lines).
	Each plot also displays an inset that is a zoomed plot of the
	cumulative distribution between 80\% and 100\% and the time
	duration between 10 hours and 1000 hours.
	%This is to clearly show the long tails of the cumulative
	%distributions.
\label{fig-duration-cumul}}
\end{figure}

We also made a table to show the statistics of low opacity conditions
continues for more than 24, 50, 100, 150, and 200 hours in unit of
counts per year (Table~\ref{tab-duration}).
The aforementioned statements are quantitatively shown in this table;
the ALMA and South Pole sites have better statistics at time duration
longer than 24 hours and 50 hours at opacity conditions lower than
0.05 and 0.04.
But for the time duration longer than 100 hours at opacity conditions
lower than 0.04, the Greenland Summit Camp clearly has better
statistics.

These results are also clearly seen in Fig.~\ref{fig-duration-cumul},
which compares the cumulative distributions of the time durations
for the three sites discussed above.
Figs.~\ref{fig-duration-cumul}(a) and (b) show the cumulative
distributions of the time durations of opacities less than 0.05 and
0.04, respectively, and it is clear that the Greenland Summit Camp
(solid line) has a long tail toward the long duration of more than
a hundred hours.
The ALMA site always exhibits higher cumulative distributions than
that of the Greenland Summit Camp, and reaches 100\% around several
tens of hours.
The South Pole site shows the steepest cumulative distribution, and
reaches 100\% around a few tens of hours, much shorter than the
other two sites.
The cumulative distributions of the time durations of opacity less
than 0.03 (Fig.~\ref{fig-duration-cumul}c) display very similar
distribution between the Greenland and the ALMA site, but the
Greenland Summit Camp shows a long tail up to several tens of hours.
Again, the South Pole site is much shorter, only up to ten hours.

In summary, for the low opacity duration, the Greenland Summit Camp
is the best site to have continuous low opacity conditions.
Since the variation of opacity is directly related to the variation
of sky temperature (background), these long stable opacity conditions
will be a significant advantage for astronomical observations, which
need unusually stable sky background, such as THz observations or
wide-field submillimeter continuum observations.

\section{Estimation of Opacities and PWVs at Other Frequencies}
\label{sect-est}

\begin{figure}[t]
\plotone{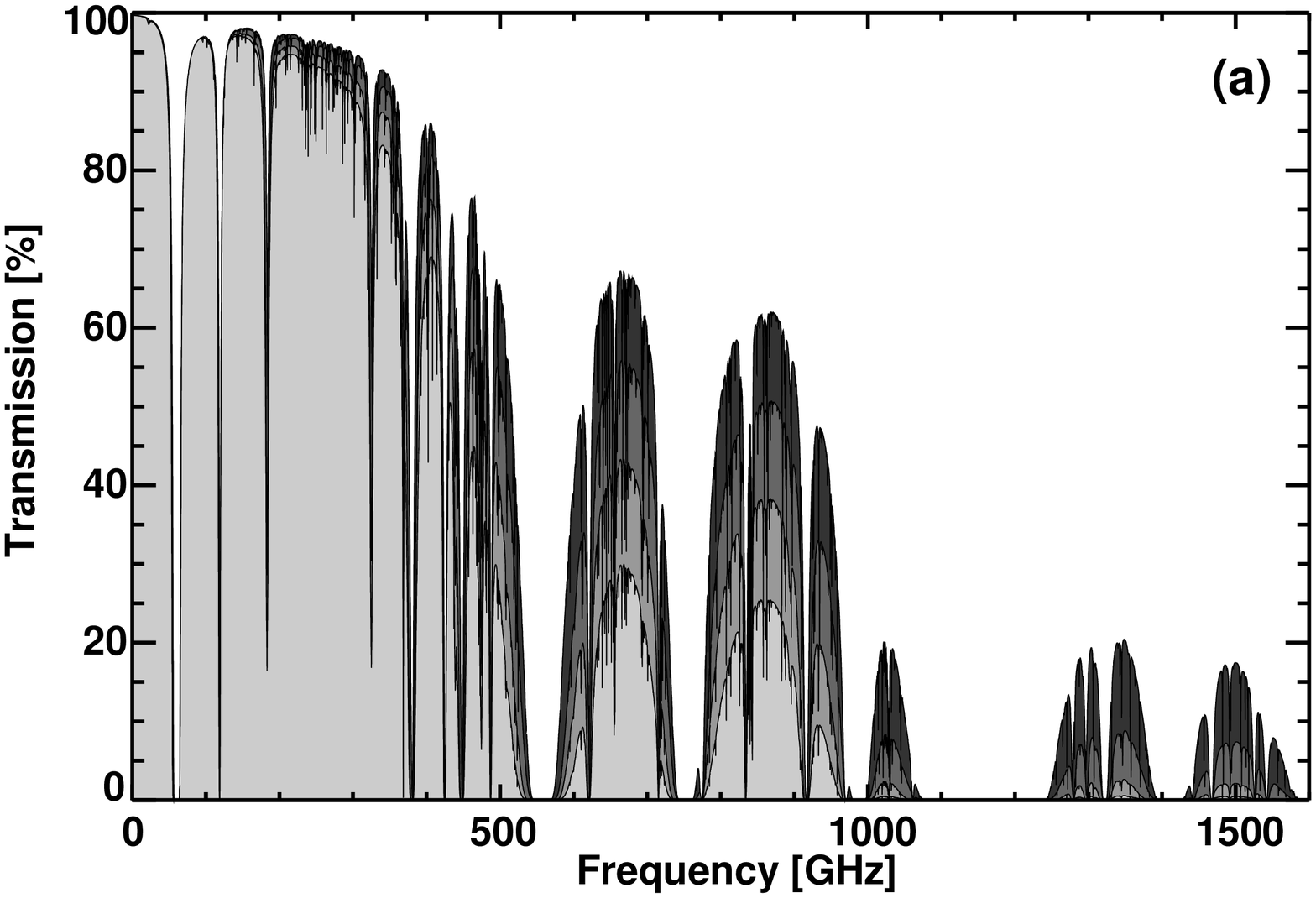}
\plotone{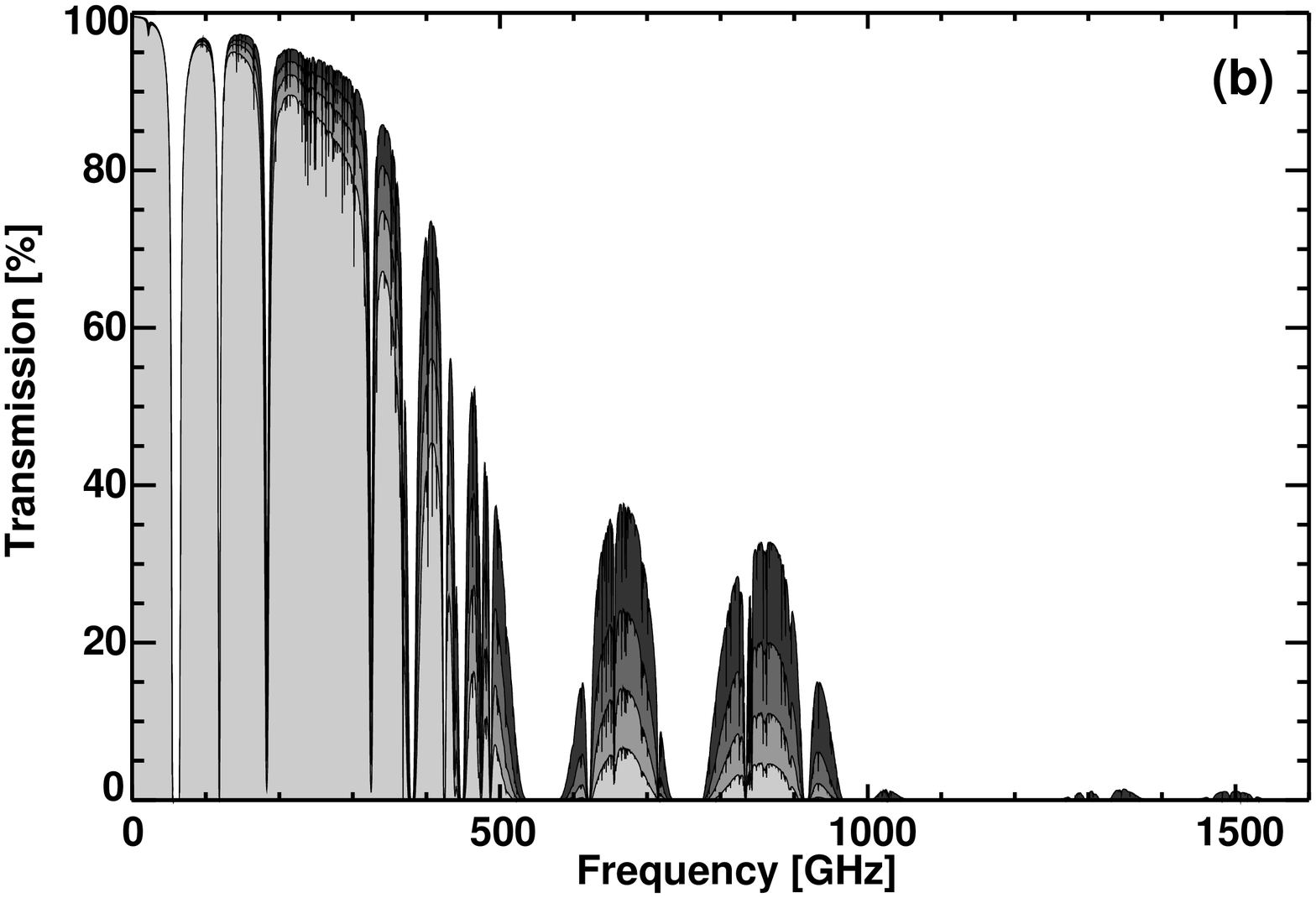}
\caption{Estimated atmospheric transmission spectra at the Greenland
	Summit Camp in (a) winter and (b) summer seasons.
	Spectra at 2\%, 10\%, 25\%, and 50\% opacity conditions are
	plotted in greyscale with darker to lighter grey.
\label{fig-est}}
\end{figure}

Using our 225 GHz opacity data and the radiative transfer program
``{\it am}'' \citep{pai14}, we estimated the atmospheric transmission
spectra between 0 GHz and 1600 GHz in both winter and summer at the
Greenland Summit Camp.
First, from the NASA Modern-Era Retrospective analysis for Research
and Applications, Version 2 \citep[MERRA-2]{rie11,mol15} reanalysis
data interpolated to the location of the Greenland Summit Camp, we
computed various percentiles of the temperature, H$_{2}$O mixing
ratio, and O$_{3}$ mixing ratio at each MERRA pressure level, for the
winter and summer months covering the same period as the radiometer
measurements.
The {\it am} model files were then constructed such that each
percentile model used the corresponding percentile temperature and
H$_{2}$O profile, whereas all models used the 50\% quartile O$_{3}$
profile.
Here we assumed that temperature and water vapor should be highly
correlated, such that it is physically meaningful to associate a
percentile profile with the corresponding percentile statistics on
each level.
%they have approximately the same percentile statistics on each level.
On the other hand, O$_{3}$ is relatively uncorrelated with either
temperature or water vapor, so that it makes sense to simply use the
median profile.

For each set of MERRA-2 percentile profiles, we found a scaling
factor on the tropospheric part of the H$_{2}$O profile, which
reproduced the corresponding 225 GHz opacity percentiles from our
measurements.
The scale factor on the median MERRA H$_{2}$O profile to match the
median 225 GHz opacity was 1.09 in winter and 1.10 in summer,
%The mean and standard deviation of these scaling factors in winter
%season was $1.10\pm0.02$,
indicating a dry bias of approximately 10\% for MERRA-2 relative to
our measurements.
%In summer season, this scaling factor is close to unity for wetter
%conditions (at the 75\% quartile), but becomes worse as conditions
%become drier (1.4 at the 2\% percentile).
%This might be an artifact of using the full six-month period data to
%represent summer, since summer is more narrowly peaked in time
%compared with winter, as shown in
%Figs.~\ref{fig-time} and \ref{fig-month}.
%Nevertheless, we think that the spectra computed with the scaled
%profiles will represent the percentile transmission spectra
%accurately.
From a radiative transfer model using the percentile profiles scaled
to our measurements, we then estimated the corresponding percentile
atmospheric transmission spectra.
%Using these model files and our 225 GHz opacity percentiles, we
%estimated the atmospheric transmission spectra.

%the National Oceanic and Atmospheric
%Administration (NOAA) 150 GHz tipping radiometer data, which is
%located just next to our radiometer, and the radiosonde profiles
%measured by NOAA at the Greenland Summit Camp, the atmospheric
%transmission spectra between 0 GHz and 1600 GHz in winter season at
%the Greenland Summit Camp have been estimated using the ``am''
%program \citep{pai14}.

The estimated atmospheric transmission spectra for the 2\%, 10\%,
25\%, and 50\% opacity conditions in winter season are plotted in
Fig.~\ref{fig-est}(a).
Note that these opacity conditions correspond to the 225 GHz opacity
of 0.031, 0.037, 0.046, and 0.060, respectively.
These spectra suggest that it is possible to observe astronomical
sources with little atmospheric attenuation (opacities $< 0.5$) most
of the winter time at the Greenland Summit Camp for frequencies lower
than 450 GHz, and half of the time for frequencies between 450 GHz
and 1000 GHz with opacities $<1.2$.
For the THz windows (1035 GHz, 1350 GHz, and 1500 GHz), 10\% of the
winter time will have an atmospheric transmission of more than 10\%.

The estimated atmospheric transmission spectra for the same opacity
percentiles as above, but in summer season are plotted in
Fig.~\ref{fig-est}(b).
Note that these opacity conditions correspond to the 225 GHz opacity
of 0.052, 0.068, 0.088, and 0.118, respectively.
These spectra suggest that it is possible to observe astronomical
sources with little atmospheric attenuation (opacities $< 0.5$) most
of the summer time for frequencies below the 380 GHz water vapor
line, and more than half of the time for the 450 GHz atmospheric
window with opacities $<1$.
For the windows between 450 GHz and 1000 GHz, 25\% of the summer
time will have an atmospheric transmission of more than 10\%.
The THz windows are totally opaque in summer.

We also calculated the relationships between 225 GHz opacity and
other submillimeter atmospheric window opacities using the above
estimates.
This is useful for estimating the opacities at higher frequencies
from the observations of the opacity at 225 GHz.
Relations between two opacities turned out to be all linear, and
therefore the linear coefficients and offsets have been calculated.
The calculated values are in Table~\ref{tab-corr}.
Opacities at the 492 GHz, 675 GHz, and 875 GHz windows are about
25 -- 30 times larger than the 225 GHz opacity, and those at the THz
windows are about $130-140$ times larger.

\begin{deluxetable}{rclc}
%\tablewidth{0pt}
\tablecaption{Calculated linear correlation between 225 GHz and submm
	atmospheric window opacities at the Greenland Summit Camp.}
\tablehead{
	\colhead{Frequency}
	& \colhead{Coefficient}
	& \colhead{Offset}
	& \colhead{Difference from} \\
	& & & \colhead{the ALMA site}
%%%%% Version 2
	}                                      %%% Difference from ALMA site
\startdata                                 %%% 1999   2000
 345 GHz &   3.80 & $-0.032$ &  $4\%$ \\   %%%  5.0%   4.1%
 410 GHz &   7.34 & $-0.064$ & $-3\%$ \\   %%%  4.1%  -2.5%
 492 GHz &  26.4  & $-0.31$  & $12\%$ \\   %%% 20.0%  11.9%
 675 GHz &  27.7  & $-0.45$  & $24\%$ \\   %%% 27.6%  23.7%
 875 GHz &  31.0  & $-0.48$  & $28\%$ \\   %%% 34.8%  28.1%
 937 GHz &  55.5  & $-0.96$  & $26\%$ \\   %%%  3.4%  26.4%
1035 GHz & 140    & $-2.7$   & $14\%$ \\   %%% 13.8%  13.8%
1350 GHz & 129    & $-2.4$   & $12\%$ \\   %%% 12.2%  12.2%
1500 GHz & 130    & $-2.3$   & $29\%$      %%% 23.8%  28.7%
\enddata
%
%%%%% Version 1
%	}                           %%% Difference from ALMA site
%\startdata                      %%% 1999   2000
% 345 GHz &   3.73 & -0.026 \\   %%%  3.0%   2.1%
% 410 GHz &   7.32 & -0.056 \\   %%%  3.8%  -2.8%
% 492 GHz &  26.8  & -0.28  \\   %%% 21.8%  13.6%
% 675 GHz &  27.4  & -0.39  \\   %%% 26.3%  22.3%
% 875 GHz &  30.1  & -0.40  \\   %%% 30.9%  24.4%
% 937 GHz &  54.4  & -0.81  \\   %%%  1.3%  23.9%
%1035 GHz & 138    & -2.3   \\   %%% 12.2%  12.2%
%1350 GHz & 122    & -1.9   \\   %%%  6.1%   6.1%
%1500 GHz & 126    & -1.9        %%% 20.0%  24.8%
%\enddata
%\tablecomments{}
\label{tab-corr}
\end{deluxetable}

These values can be compared with the measurement results at the ALMA
site \citep{mat99,mat00}:
For the 345 GHz and 410 GHz windows, there are only a few \%
differences between the Greenland Summit Camp and the ALMA site, but
for higher frequencies, the Greenland Summit Camp is about $10-15$\%
worse than the ALMA site for the 492 GHz, 1035 GHz, and 1350 GHz
windows, and about 25\% worse for the 675 GHz, 875 GHz, 937 GHz, and
1500 GHz windows (Table~\ref{tab-corr}).
This is due to the altitude difference.
The altitude of the Greenland Summit Camp is only 3200 m, much lower
than that of the ALMA site at 5000 m, and this difference increases
the opacity in the pressure-broadened wings of saturated H$_{2}$O
lines at high frequencies.
In addition, the lower altitude introduces larger dry air continuum
absorption at the Greenland Summit Camp; the dry air continuum
absorption increases at higher frequencies, up to the middle of the
N$_2$-N$_2$ collision-induced absorption band near 3 THz
\citep{par01a,par01b,pai14}.
%\citep{par01a,par01b,pai14}, and therefore it affects the opacities
%at higher frequencies much more than at lower frequencies.

%In summary, lower frequency ($<450$ GHz) observations in Greenland
%will be almost the same condition as at the ALMA site, but
%higher frequency ($>450$ GHz) observations will be 10\% -- 25\% more
%difficult.
%Millieter and submm transmission up to 1000 GHz are good, and
%therefore the observations at these frequencies are well promised.
%Transmission between 1000 GHz and 1600 GHz is not high, but it is
%possible.

The ``{\it am}'' program also estimates the PWVs together with the
transmission spectra, and we present those in Table~\ref{tab-pwv}.
Using these values, it is possible to derive the relation between the
225 GHz opacity and PWV at the Greenland Summit Camp, which turned
out to be
\begin{equation}
\tau_{\rm 225 GHz} = 0.048 \times {\rm PWV~[mm]} + 0.022,
\end{equation}
%for winter and
%\begin{equation}
%\tau_{\rm 225 GHz} = 0.046 \times PWV + 0.023
%\end{equation}
%for summer,
where $\tau_{\rm 225 GHz}$ is the 225 GHz opacity.
With this equation, together with the relationship between the
225 GHz opacity and that of other frequencies (Table~\ref{tab-corr}),
it is also possible to compare various atmospheric window opacities
and PWV.

\begin{deluxetable}{rcccc}
\tablecaption{Estimated PWV for winter and summer seasons.}
%\begin{deluxetable}{rcccccc}
%\tablecaption{Estimated PWV for annual, winter, and summer.}
\tablehead{
%	& \multicolumn{2}{c}{Annual\tablenotemark{a}}
	& \multicolumn{2}{c}{Winter}
	& \multicolumn{2}{c}{Summer} \\
	\colhead{Percentile}
%	& \colhead{225 GHz}
%	& \colhead{PWV}
	& \colhead{225 GHz}
	& \colhead{PWV}
	& \colhead{225 GHz}
	& \colhead{PWV} \\
%	& \colhead{Opacity}
%	& \colhead{[mm]}
	& \colhead{Opacity}
	& \colhead{[mm]}
	& \colhead{Opacity}
	& \colhead{[mm]}
	}
\startdata
 2\% & 0.0311 & 0.186 & 0.0516 & 0.623 \\
10\% & 0.0375 & 0.321 & 0.0679 & 0.965 \\
25\% & 0.0464 & 0.506 & 0.0885 & 1.396 \\
50\% & 0.0602 & 0.790 & 0.1178 & 2.054 \\
75\% & 0.0801 & 1.205 & 0.1586 & 2.942
%1\% & 0.0303 & 0.171 & 0.0292 & 0.147 & 0.0453 & 0.498 \\
% 2\% & 0.0325 & 0.216 & 0.0311 & 0.186 & 0.0516 & 0.623 \\
%3\% & 0.0337 & 0.243 & 0.0323 & 0.211 & 0.0556 & 0.705 \\
%5\% & 0.0364 & 0.299 & 0.0338 & 0.244 & 0.0606 & 0.808 \\
%10\% & 0.0425 & 0.425 & 0.0375 & 0.321 & 0.0679 & 0.965 \\
%25\% & 0.0578 & 0.740 & 0.0464 & 0.506 & 0.0885 & 1.396 \\
%50\% & 0.0836 & 1.284 & 0.0602 & 0.790 & 0.1178 & 2.054 \\
%75\% & 0.1258 & 2.225 & 0.0801 & 1.205 & 0.1586 & 2.942
\enddata
%\tablenotetext{a}{Annual values are derived using the 3-year subset
%	from October 2011 to September 2014.}
\label{tab-pwv}
\end{deluxetable}

The 3-year (2008 -- 2010) PWV statistics and the atmospheric
transmission spectra of the summit of Greenland have also been
derived by \citet{tre12} using the water vapor product from the
Infrared Atmospheric Sounding Interferometer (IASI) instrument on the
Meteorological Operation (MetOp)-A satellite.
They found a median annual PWV of 0.94 mm over the summit of
Greenland.
Our radiometer-derived median PWV for the 3-year subset of our data
from October 2011 to September 2014 is 1.28 mm.
% (Table~\ref{tab-pwv}).
To connect these two periods, we note that the MERRA-2 median PWV for
January 2008 -- December 2010 is 1.27 mm, whereas for October 2011 --
September 2014 it is 1.16 mm.
If we assume the MERRA-2 dry bias of approximately 10\% is consistent
between these periods, then the implied median PWV for 2008 -- 2010
is 1.40 mm.
%This would suggest that the PWV for Summit in \citet{tre12} has dry
%bias of approximately 33\%.
This would suggest that the PWV for Summit in \citet{tre12} is
systematically lower than the actual PWV by approximately 33\%.

There are two reasons why a single satellite data set could produce
significantly biased PWV statistics.
First, although IASI is able to retrieve surface temperature, the
water vapor retrieval accuracy in the lower troposphere necessarily
suffers from a lack of thermal contrast with the surface
\citep{wul15}.
The second reason is temporal sampling bias.
Typically satellite sounders are in sun-synchronous orbits that pass
over a given point on the Earth's surface at the same pair of local
times each day; these are 9:30 and 21:30 in the case of IASI
\citep{hil12}.
In comparison, a reanalysis such as MERRA-2 can be expected to
produce more realistic PWV statistics because it assimilates multiple
satellite, surface, and upper air measurements using a model with
realistic dynamics.

\section{Summary}
\label{sect-sum}

We present the 3.5-yr monitoring of the 225 GHz opacity at the Summit
of the Greenland ice sheet using a tipping radiometer.

Opacity variations clearly show a seasonal variation between winter
($\approx$ nighttime) and summer ($\approx$ daytime), but no diurnal
variation, with the opacity in winter being about half of that in
summer.
This is similar to the opacity variations at the South Pole site due
to the polar conditions, but the absolute opacity value is about
10\%-30\% higher than that of the South Pole site in winter, and about
double in summer.
In contrast, the ALMA site shows clear seasonal and diurnal
variations due to mid-latitude conditions; the opacity at the ALMA
site is up to 20\% better than at the Greenland site in winter, but
up to 40\% worse in summer.

The estimated atmospheric transmission spectra suggest that most of
the winter time is useable for astronomical observations at
frequencies lower than 450 GHz, half of the time is useable for
frequencies between 450 GHz and 1000 GHz, and 10\% of the time is
useable for the THz atmospheric windows.
Most of summer time is useable at frequencies lower than 380 GHz, and
half of the summer time is useful at the 450 GHz atmospheric window.
The linear correlations between 225 GHz and submillimeter atmospheric
window opacities are derived, and opacities at the 492 GHz, 675 GHz,
and 875 GHz windows are about 25 -- 30 times larger, while those at
THz windows are about $130-140$ times larger than the opacity at
225 GHz.
These opacities are up to 25\% higher than those at the ALMA site,
which is due to the altitude difference between the Greenland Summit
Camp (3200 m) and the ALMA site (5000 m).
%as explained above.
%and therefore the difference of the amount
%of dry atmosphere above the sites.

The biggest advantage of the opacity conditions at the Greenland
Summit Camp is the long time durations of low opacity.
At the Greenland Summit Camp, opacities lower than 0.04 or 0.05 can
continue for more than a hundred hours occasionally, and opacities
lower than 0.03 can continue for more than several tens of hours in
some cases.
%This is much longer than the opacity conditions at the ALMA or
%the South Pole sites,
In case of opacity lower than 0.04, the Greenland Summit Camp is the
only site that can continue this condition for more than one hundred
hours, indicating that the Greenland site is suitable for
observations that need stable opacity conditions for a long time.

\acknowledgements
We would like to thank CPS for providing liquid nitrogen yearly,
which was used for the calibration of the radiometer's detector, and
to the ICECAPS technicians who assisted us in a timely fashion when
needed.
We also would like to thank Simon J. E. Radford for providing us with
the opacity data of the ALMA and the South Pole sites.
The MERRA-2 data used in this study have been provided by the Global
Modeling and Assimilation Office (GMAO) at NASA Goddard Space Flight
Center.
We are also grateful to the anonymous referee for helpful comments.
SM is supported by the National Science Council (NSC) and the
Ministry of Science and Technology (MoST) of Taiwan,
NSC 100-2112-M-001-006-MY3 and MoST 103-2112-M-001-032-MY3.

\end{document}